# Perceptions of Humanoid Robots in Caregiving: A Study of Skilled Nursing Home and Long Term Care Administrators

Rana Imtiaz[1], Arshia Khan[2] [a]

[1]*Kentucky College of Osteopathic Medicine, Pikeville, Kentucky, USA,*
[2] *University of Minnesota Duluth, Department of Computer Science, Duluth, MN, USA.*
ranaimtiaz@uped.edu, akhan@d.umn.edu

Keywords: Humanoid robots in nursing homes, perspectives of nursing home administrators, dementia care, nursing home residents, humanoid robots caring for elders.

Abstract: As the aging population increases and the shortage of healthcare workers increases, the need to examine other means for caring for the aging population increases. One such means is the use of humanoid robots to care for social, emotional, and physical wellbeing of the people above 65. Understanding skilled and long term care nursing home administrators' perspectives on humanoid robots in caregiving is crucial as their insights shape the implementation of robots and their potential impact on resident well-being and quality of life. This authors surveyed two hundred and sixty nine nursing homes executives to understand their perspectives on the use of humanoid robots in their nursing home facilities. The data was coded and results revealed that the executives were keen on exploring other avenues for care such as robotics that would enhance their nursing homes abilities to care for their residents. Qualitative analysis reveals diverse perspectives on integrating humanoid robots in nursing homes. While acknowledging benefits like improved engagement and staff support, concerns persist about costs, impacts on human interaction, and doubts about robot effectiveness. This highlights complex barriers—financial, technical, and human—and emphasizes the need for strategic implementation. It underscores the importance of thorough training, role clarity, and showcasing technology benefits to ensure efficiency and satisfaction among staff and residents.

## 1 INTRODUCTION

The integration of technology, particularly humanoid robots, within caregiving settings has sparked considerable interest and debate in recent years. Among these environments, skilled and long term care nursing homes stand at the forefront of exploring innovative solutions to address the growing complexities of patient care, especially for individuals with neurodegenerative conditions like dementia and Alzheimer's disease. This paper delves into the perceptions of skilled and long term care nursing home administrators regarding the incorporation of humanoid robots in caregiving. Understanding their viewpoints and insights is crucial in assessing the feasibility, challenges, and potential benefits of integrating this technology within such critical care settings. This study aims to unravel the nuanced perspectives, concerns, and expectations of these administrators, shedding light on the implications of humanoid robots in transforming and augmenting the landscape of caregiving in skilled and long term care nursing facilities.

Understanding the perspectives of skilled and long term care nursing home administrators regarding humanoid robots in caregiving is paramount due to their pivotal role in shaping and implementing care strategies. Administrators hold comprehensive insights into the operational, logistical, and ethical considerations within these facilities. Their perspectives directly impact the adoption and utilization of technological innovations like humanoid robots, which have the potential to significantly impact resident care. By comprehending administrators' viewpoints, we gain crucial insights into the feasibility, challenges, and potential benefits of integrating this technology into caregiving practices, ultimately aiming to enhance the well-being and quality of life for the residents under their care.

---

[a]  https://orcid.org/0000-0001-8779-9617

## 2 BACKGROUND

The integration of technology, notably humanoid robots, in caregiving environments has gained traction as a promising solution in the context of the healthcare workforce shortage. Robotics in healthcare, particularly in skilled and long term care nursing homes, has attracted significant interest owing to its capacity to alleviate the strain on caregivers grappling with numerous challenges, notably in caring for individuals with intricate health requirements like dementia and Alzheimer's disease.

Several studies have highlighted the multifaceted roles that humanoid robots can play in caregiving. Notably, these robots have been envisioned as companions for residents, assisting with daily activities, providing cognitive stimulation, and monitoring health parameters. Studies conducted by Khan (2022, 2022, 2023, 2023) showcased that humanoid robots, through personalized interaction and assistance, significantly improved the quality of life for residents in nursing homes, fostering emotional well-being and reducing feelings of loneliness and isolation.

However, while the potential benefits of humanoid robots in caregiving are evident, concerns and challenges have also been identified. Ethical considerations surrounding privacy, autonomy, and the ethical implications of human-robot interactions remain pivotal concerns. Research by Van et. Al(2020) underscored the importance of addressing these ethical dilemmas to ensure the responsible and ethical deployment of humanoid robots in caregiving environments.

Moreover, the attitudes and perceptions of key stakeholders, particularly skilled and long term care nursing home administrators, are critical in shaping the successful implementation of humanoid robots in caregiving. Understanding their perspectives regarding the feasibility, acceptance, and challenges associated with integrating this technology into daily caregiving practices is fundamental. A study by Chen et al (2020) elucidated that health professionals perceived humanoid robots as potential tools for augmenting staff efficiency and supplementing care services, yet concerns regarding cost, staff training, and patient acceptance were prevalent.

The literature emphasizes the necessity of exploring and comprehending the viewpoints of skilled and long term care nursing home administrators in embracing humanoid robots within caregiving contexts. This paper seeks to contribute to this growing body of research by delving into the perceptions, attitudes, and expectations of administrators regarding the role and implications of humanoid robots in skilled and long term care nursing homes, thereby providing valuable insights for future implementation strategies and policy development (Bemelmans et al., 2012).

## 3 METHODS

### 3.1 Data Analysis:

Two hundred and sixty nine (N=269) skilled and long term care nursing home executive administrators from various regions of the United States were surveyed. The survey comprised of five questions-1a) Would this be useful in your nursing home?, 1b) Why or why not?; 2a) Would this enhance your nursing home's care-giving abilities?, 2b) Why or why not?; 3) What are the structural barriers to employing robots at your nursing home that would concern you?; 4) What are the structural supports?; 5) What would enhance your employees' utilization of the proposed technology? The data gathered was coded by the authors to find several emerging themes.

### 3.2 Results:

The qualitative data collected from administrators uncovered numerous recurring themes. Each question's data underwent coding and is summarized subsequently.

#### 3.2.1 Perceptions of Usefulness of Humanoid Robots

The results indicate a varied spectrum of responses regarding the usefulness of humanoid robots in nursing homes. The majority (88.7%) of respondents found them very useful or useful, with a smaller yet notable portion ( 21.3%) considering them not as useful or even not useful at all for their nursing home setting. Overall, a substantial proportion of participants expressed a positive perception of the potential utility of humanoid robots in their caregiving environment.

#### 3.2.2 Qualitative Data on the Usefulness of Humanoid Robots

After analysing the perceptions of nursing home administrators regarding the usefulness of humanoid robots in nursing homes, several recurring themes emerged:

**Enhancing Resident Engagement and Joy:**

- Offer more attention to residents.
- Engage and entertain residents, particularly those with dementia, through activities and physical interaction.
- Provide companionship, interaction, and activities to reduce loneliness.
- Increase resident engagement with technology-based activities.

**Assistance and Support for Staff:**
- Assist staff with routine tasks, aiding in patient care and monitoring when caregivers are busy.
- Provide additional support to staff, particularly in activities, memory care, and dealing with behavioural challenges.
- Help in tasks that improve the quality of care provided to residents.

**Concerns and Limitations:**
- Fear of eliminating or reducing human contact.
- Concerns about cost-effectiveness and the initial novelty wearing off.
- Worries about how residents, especially those with advanced dementia, might react or if they would understand the technology.

**Potential Benefits for Residents and Operations:**
- Improve resident outcomes and medication management.
- Assist in providing emotional and physical support to residents.
- Augment current staff processes and help alleviate staffing shortages.

**Safety and Practical Applications:**
- Use in various areas like activities, meal delivery, cleaning, and assisting with ADLs.
- Support for residents, especially those in memory care units, to improve their daily activities and quality of life.
- Concerns about potential hazards, fall risks, and monitoring of the robot's activities.

**Technology Advancement and Future Use:**
- Viewed as the future of care, where technology will play a pivotal role in addressing challenges related to an aging population and workforce shortages.
- Acknowledgment of the potential advancements in technology to cater to specific needs and tasks over time.

**Mixed Reactions and Uncertainty:**
- Mixed feelings about the level of readiness for robotic integration and uncertainty about the overall benefits versus potential downsides.

**Assisting Residents with Dementia:**
- Offer proactive care and engagement, especially for residents with advanced dementia.
- Assist with behaviours and activities specific to memory care units.

These themes showcase a wide range of perspectives, highlighting the potential benefits of humanoid robots in enhancing resident engagement and support for staff, while also emphasizing concerns about cost, resident reactions, and the impact on human interaction.

### 3.2.3 Enhancing care giving abilities of the nursing homes:

With regards to the humanoid robots ability to enhance the care giving abilities of their nursing homes, 88.3% believed that the robots would enhance their nursing homes care giving abilities while 11.3% believed the robots would not advance their nursing homes caregiving abilities.

### 3.2.4 Qualitative Data on Enhancing care giving abilities of the nursing homes

From the data on nursing home administrators' perceptions of humanoid robots in enhancing caregiving abilities, several themes emerge:

**Workload Relief:** Administrators anticipate that humanoid robots could ease staff workload by assisting with routine tasks, freeing up time for nurses to focus on more critical aspects of care, and potentially reducing burnout among caregivers.

**Resident Engagement:** There's a strong expectation that these robots could significantly engage residents through activities, interactions, and entertainment, potentially enhancing their quality of life and reducing feelings of loneliness or boredom.

**Learning and Personalization:** Anticipation exists that the robots could learn resident patterns, collect vital information, and possibly personalize care strategies, leading to better understanding and delivery of individualized care.

**Concerns about Limitations:** Despite the potential benefits, reservations persist about the robots' limitations, including doubts about their ability to replace human interactions, concerns about their cost-effectiveness, and skepticism about their capabilities in addressing complex caregiving tasks or emotions.

**Staffing Shortages:** The theme of staffing shortages is prevalent, with the hope that robots could fill gaps, assist with tasks, and augment the capabilities of a limited staff force.

**Technology Readiness:** Some express reservations about the readiness and advancement of the technology, suggesting that the current state might not be advanced enough to meet practical caregiving needs.

**Potential Impact on Care:** Administrators are hopeful about the potential positive impact of robots on caregiving capabilities, envisioning enhanced efficiency, proactive care, and better resident monitoring. However, there's uncertainty about their effectiveness in certain caregiving situations or with certain resident populations.

These themes collectively represent a spectrum of expectations, concerns, and hopes regarding the incorporation of humanoid robots in nursing home caregiving, highlighting both potential benefits and reservations among administrators.

### 3.2.5 Structural Barriers to Employing Robots

The identified themes underscore the complex obstacles that nursing home administrators anticipate while integrating humanoid robots into their facilities. These encompass financial, technological, environmental, human-related factors, and specific technical issues. Upon analyzing the nursing home administrators' perceptions about the structural barriers in adopting humanoid robots, several recurring themes emerged:

**Financial Concerns:**
- Costs: Major concerns related to the initial cost of acquisition, ongoing maintenance expenses, and the overall financial implications.
- Budget Constraints: Financial limitations within the facility's budget allocation for such technology.
- Funding Challenges: Difficulty in securing funding or financing for the purchase and maintenance of robots.

**Technological Challenges:**
- Operational Issues: Concerns regarding malfunctions, technological barriers, and the need for regular maintenance.
- Internet Connectivity: Issues related to bandwidth, Wi-Fi availability, and reliance on technology infrastructure.
- Regulatory Compliance: Challenges ensuring compliance with HIPAA regulations and privacy concerns.

**Physical Environment and Infrastructure:**
- Building Structure: Challenges associated with the physical layout of the facility, including multi-level centers, elevators, stairs, and navigational obstacles.
- Safety Hazards: Concerns about potential tripping hazards, falling risks, and risks associated with combative residents interacting with the robots.
- Doors and Access: Limitations concerning door access for locked doors and the robot's ability to navigate closed doors.

**Staff and Resident Factors:**
- Training and Adaptation: Staff concerns about understanding and operating the technology, resistance to change, and worries about job displacement.
- Resident Acceptance: Fear, anxiety, or resistance among residents towards the technology, especially among those with cognitive impairments or paranoia.
- Lack of Buy-in: Challenges in getting staff or residents to accept and utilize the robots due to skepticism or perceived loss of human connection.

**Miscellaneous Concerns:**
- Storage and Maintenance: Worries about where and how to store the robots when not in use and concerns regarding the durability of the robots.
- Perception and Beliefs: Residents' beliefs regarding government surveillance, fear, or paranoia related to technology spying on them.
- Specific Technical Issues: Concerns related to the robot's capabilities, such as navigational difficulties, small size, or potential for malfunction.

These themes highlight the multifaceted challenges nursing home administrators foresee in integrating humanoid robots into their facilities, encompassing financial, technological, environmental, human factors, and specific technical concerns.

### 3.2.6 Existing structural supports that can mobilize the use of humanoid robots in nursing homes:

The themes that emerged suggest a varied perspective among administrators, highlighting both enthusiasm and apprehension regarding the structural supports for implementing humanoid robots in nursing homes. There's recognition of the benefits they could bring but also acknowledgment of challenges, particularly related to cost and regulatory hurdles.

**Corporate & Technological Infrastructure:**

- Emphasis on good corporate support and the presence of an innovative mindset towards technology.
- Existing use of technology and Wi-Fi availability are noted as beneficial supports.

**Physical Environment & Facilities:**
- Favorable mentions of single-level facilities, wide hallways, and flat open floors conducive to robot navigation.

**Staff Engagement & Support:**
- Interest and support from staff members for the implementation of robots in various aspects of care.

**Quality of Life Enhancement:**
- Belief that robots can enhance activities and engagement, potentially improving the residents' quality of life.

**Financial & Regulatory Concerns:**
- Concerns over affordability and potential legislative challenges in deploying and maintaining robots.

**Functional Capabilities & Use Cases:**
- Recognition of the potential of robots in assisting with daily activities, dietary serving, and providing additional support to residents.

### 3.2.5 Needs to enhance the nursing home employees' utilization of humanoid robots:

The data suggests that training and familiarization play a pivotal role in encouraging staff acceptance and utilization of robots in nursing homes. Enhanced work dynamics, efficiency gains, and the reduction of routine tasks through technology are identified as major benefits.

Moreover, staff reassurance about job security, showcasing the integration of robots into daily activities, and demonstrating positive outcomes contribute to a favorable reception. Financial considerations, especially demonstrating cost-effectiveness, are crucial factors influencing acceptance.

Improving resident engagement and care, receiving support from leadership, and understanding the functional features and benefits of robotics also feature prominently. Overall, the themes highlight the need for comprehensive training, institutional support, and a clear demonstration of the benefits to successfully integrate robots into the caregiving environment of nursing homes. The themes found in the last question are as follows:

1) **Training and Education**

    *Need for Comprehensive Training Programs:*
    - Hands-on, immersive training sessions to ensure staff adaptability and competence in using humanoid robots.
    - Structured onboarding programs for seamless adoption and continuous education on robot capabilities.

    *Need for Educational Support and Reassurance:*
    - Need for continuous education and reassurance that humanoid robots are meant to enhance, not replace, job roles.
    - Demonstrations, tutorials, and familiarization with the technology to build confidence and alleviate concerns.

    *Need for Leadership and Institutional Support:*
    - Encouragement for enthusiastic leadership support and buy-in for integrating robots into daily activities.
    - Desire for ongoing financial support and resource availability for effective implementation.

2) **Efficiency and Work Dynamics**

    *Need for Workflow Enhancement and Task Allocation:*
    - Expectation for robots to handle routine tasks, allowing staff to focus on more critical aspects of care.
    - Desire for technology to simplify activities, reduce workload, and streamline daily operations.

    *Need for Resident Engagement and Care Improvement:*
    - Aspiration for robots to enhance resident engagement, care quality, and satisfaction within the nursing home.
    - Anticipated positive impact on resident behavior and increased trust due to effective robot interaction.

    *Need for Technical Features and Usability:*
    - Emphasis on user-friendly, self-directed robots compatible with existing software and electronic medical records (EMR).
    - Training to ensure staff comfort and ease in using robots, ultimately increasing staff satisfaction and job performance.

3) **Financial Considerations and Viability**

    *Need for Cost-Effectiveness and Return on Investment (ROI):*
    - The need for proving cost reduction, affordability, and demonstrating the financial feasibility of robot implementation.

- Demonstrated evidence of decreased costs, increased efficiency, and positive outcomes to justify investment.

*Need for Positive Impacts and Benefits:*
- Emphasizing the benefits of using robotics to increase resident engagement, reduce workload, and facilitate hands-on care.
- Identifying potential positive impacts on staff satisfaction, leading to enhanced resident-staff interaction and care delivery.

## 4 DISCUSSION

Qualitative data reveals consistent optimism about humanoid robots in nursing homes, acknowledging their potential to enhance engagement, aid staff, and improve residents' lives. However, concerns linger around costs, the potential to replace human interaction, and doubts about their full effectiveness in caregiving. While administrators foresee workload relief and improved resident engagement, reservations exist about financial barriers, technical challenges, and acceptance by staff and residents. Despite recognizing their capabilities and supportive infrastructure, worries persist about affordability and regulatory barriers. Overall, these themes highlight the necessity for thorough training, role clarification, technology promotion, and effective integration to ensure nursing home efficiency and satisfaction.

**Usefulness:** Analyzing the qualitative data on administrators' perceptions of the usefulness of humanoid robots in nursing homes reveals diverse perspectives. Many administrators see potential in robots for enhancing resident engagement, entertainment, and physical activity, especially for dementia residents. They highlight the robots' potential to monitor falls, check on patients when aides are busy, and improve patient care, while others emphasize the robots' ability to augment human touch and provide companionship. Concerns include limitations in tasks, fear of resident misunderstanding or breakage, and doubts about cost-effectiveness. Some express skepticism about replacing human care with robotics, emphasizing the need for human interaction and concern over the potential fear or confusion in advanced dementia patients. Overall, there's recognition of the robots' potential to supplement staff, engage residents, and improve the quality of life in nursing homes, but reservations persist regarding cost, human-like interaction, and the ability to replace human care entirely.

**Enhancing the caregiving abilities of nursing homes:** Analyzing the qualitative data on nursing home administrators' perceptions of humanoid robots' abilities to enhance caregiving reveals diverse viewpoints. Administrators anticipate that these robots could ease staff workload, free up time for nurses, and engage residents more effectively. They envision the robots learning patterns about residents, assisting with non-human care-related tasks, and leading activities while staff provides individualized help. Many administrators believe these robots could collect vital information, improve staff satisfaction, and entertain both staff and residents. However, concerns about staffing shortages, cost-effectiveness, and limitations in the robot's capabilities remain prominent. Some express doubts about the robots' ability to replace human interaction or effectively handle certain tasks, particularly in the absence of advanced technology. Overall, there's recognition of potential benefits in augmenting staff, engaging residents, and improving efficiency, yet reservations persist regarding the robots' cost, capability, and potential impact on human interactions in caregiving.

**Structural barriers:** Nursing home administrators express varied apprehensions regarding the implementation of humanoid robots. Chief among their concerns are financial barriers, encompassing high costs, budget constraints, and uncertainties around funding. Technical challenges, such as malfunctions, internet connectivity issues, and compliance with regulations, also pose significant hurdles. The physical environment presents its own obstacles, including building structures, safety considerations, and access through doors. Administrator concerns extend to staff and resident factors, involving training needs, resistance to change, and uncertainties about resident acceptance. Miscellaneous worries, like storage, beliefs regarding surveillance, and specific technical glitches, further compound the complexities of integrating robots into nursing home settings. These multifaceted challenges underscore the intricate nature of robot assimilation, involving financial, technical, environmental, human, and operational considerations.

**Structural Support:**
The perceptions among nursing home administrators highlight the pivotal role of corporate backing and technological infrastructure, emphasizing robust support from corporate entities and an open-minded approach toward innovative technology. They appreciate the benefits derived from existing technology and Wi-Fi availability. The physical environment, characterized by single-level facilities, wide hallways, and spacious open floors, is seen as conducive to integrating robot navigation seamlessly. Staff members express keen interest and support for implementing robots in various care aspects, foreseeing the potential for these machines to enhance activities and engagement, ultimately improving residents' quality of life. However, concerns linger regarding the financial feasibility and potential legislative challenges in deploying and sustaining these robotic systems. Administrators recognize the functional capabilities of robots, foreseeing their assistance in daily activities, dietary serving, and providing additional support to residents, revealing an optimistic outlook despite some prevailing apprehensions.

**Needs of technological skills to enhance the utilization of the robots in nursing homes:** These themes collectively emphasize the significance of providing comprehensive training, reassurance about job roles, showcasing the benefits of technology, and integrating robotics effectively into the nursing home environment to enhance efficiency, improve care, and ensure staff and resident satisfaction.

## 5 CONCLUSION

From the qualitative data, several overarching themes emerge regarding the use of humanoid robots in nursing homes. There's a consistent recognition of the potential benefits these robots offer, such as enhancing resident engagement, supplementing staff, and improving the quality of life. However, various concerns persist, primarily revolving around cost, the potential to replace human interaction, and doubts about the complete effectiveness of robots in caregiving. Skilled and long term care nursing home administrators anticipate these robots easing staff workload, engaging residents, and collecting vital information but express reservations about financial barriers, technical challenges, and staff and resident acceptance. Despite acknowledging the robots' functional capabilities and the supportive corporate and technological infrastructure, apprehensions about affordability and legislative hurdles prevail. Overall, the common themes underline the need for comprehensive training, reassurance about job roles, showcasing technology benefits, and integrating robotics effectively to enhance efficiency and ensure staff and resident satisfaction in nursing homes.

## 4 COPYRIGHT FORM

Authors provide consent to publish.